\documentclass[twocolumn,showpacs,preprintnumbers]{revtex4}

\usepackage{amsmath}
\usepackage{amsfonts}
\usepackage{amsthm}
\usepackage{amssymb}
\usepackage{graphicx}
\usepackage{mathrsfs}
\usepackage{latexsym}
\usepackage{graphics}
\usepackage{cases}
\usepackage{hyperref}

\def\mathbi#1{\textbf{\em #1}}
\newcommand{\beq}{\begin{equation}}
\newcommand{\eeq}{\end{equation}}
\renewcommand{\d}{{\rm d}}
\newcommand{\del}{\mbox{\boldmath$\nabla$}}
\newcommand{\x}{{\mathbi x}}

\renewcommand{\v}{{\mathbi v}}
\newcommand{\g}{{\mathbi g}}
\newcommand{\h}{{\mathbi h}}
\newcommand{\e}{\hat{\mathbf{e}}}
\newcommand{\F}{{\mathbi F}}
\newcommand{\R}{{\mathbi R}}
\renewcommand{\S}{{\mathbi S}}
\newcommand{\I}{{\mathbi I}}

\newcommand{\Vvaporjetobs}{{V_{\rm jet}^\ast}}
\newcommand{\Vvaporjetobsmax}{{V_{\rm jet}^{\ast\rm\,max}}}
\newcommand{\Vvaporjetcor}{V_{\rm jet}}

\newcommand{\ejet}{\epsilon_{\rm jet}}

\newcommand{\rmax}{R_0}
\newcommand{\rreb}{R_1}
\newcommand{\deltap}{\Delta p}

\newcommand{\fig}[1]{Fig.~\ref{fig_#1}}
\newcommand{\eq}[1]{Eq.~(\ref{eq_#1})}
\newcommand{\eqp}[1]{Eq.~\ref{eq_#1}}

\begin{document}

\title{A Universal Scaling Law for Jets of Collapsing Bubbles}
\author{D. Obreschkow$^1$}
\author{M. Tinguely$^1$}
\author{N. Dorsaz$^2$}
\author{P. Kobel$^3$}
\author{A. de Bosset$^1$}
\author{M. Farhat$^1$}

\affiliation{
$^1\,$EPFL, Laboratoire des Machines Hydrauliques, 1007 Lausanne, Switzerland\\
$^2\,$Department of Chemistry, University of Cambridge, Cambridge CB2 1EW, UK\\
$^3\,$Max-Planck-Institut f\"ur Sonnensystemforschung, 37191 Katlenburg-Lindau, Germany}

\pacs{47.55.dp,47.55.dd,43.25.Yw}

\date{\today}

\preprint{Physical Review Letters, accepted, 2011}

\begin{abstract}
Cavitation bubbles collapsing and rebounding in a pressure gradient $\del p$ form a ``micro-jet'' enveloped by a ``vapor jet''. This letter presents unprecedented observations of the vapor jets formed in a uniform gravity-induced $\del p$, modulated aboard parabolic flights. The data uncovers that the normalized jet volume is independent of the liquid density and viscosity and proportional to $\zeta\equiv|\del p|\rmax/\Delta p$, where $\rmax$ the maximal bubble radius and $\Delta p$ is the driving pressure. A derivation inspired by ``Kelvin-Blake'' considerations confirms this law and reveals its negligible dependence of surface tension. We further conjecture that the jet only pierces the bubble boundary if $\zeta\gtrsim4\cdot10^{-4}$.
\end{abstract}

\maketitle

\begin{figure}[t!]
	\centering 
  	\includegraphics[width=\columnwidth]{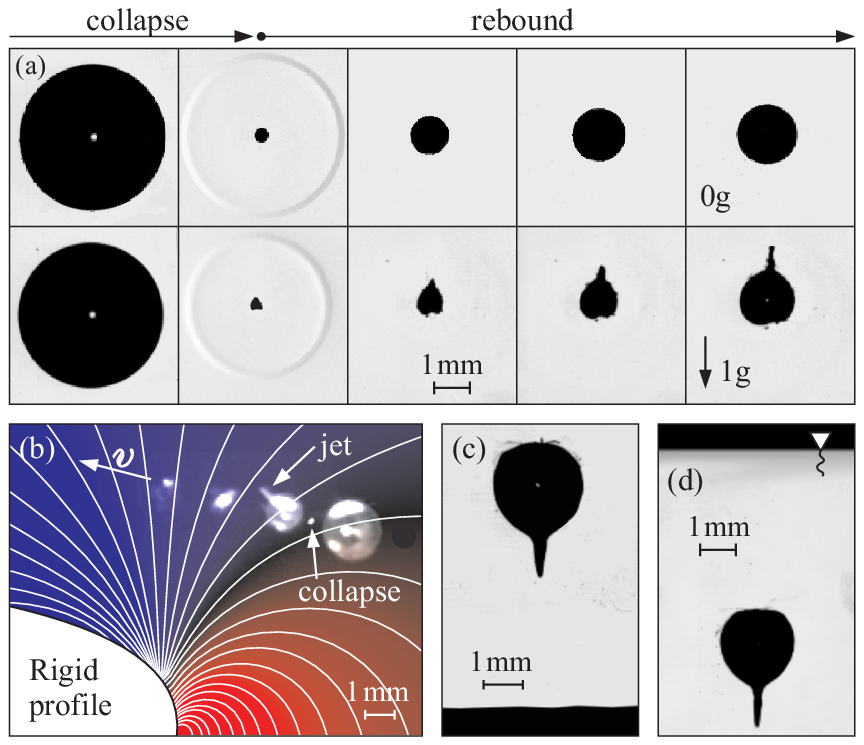}
  	\caption{Observations of the vapor jets directed against $\del p$ during the rebound of cavitation bubbles. (a--d) match the cases of Eqs.~(\ref{eq_delp_all}a--d): (a) [\href{http://quantumholism.com/pdf/featured_movie.wmv}{video on-line}] Collapse and rebound of a bubble ($R_0\approx4\rm\,mm$, $\Delta p\approx15\rm\,kPa$) in `0g' (upper) and `1g' (lower); note the shock at the collapse. (b) Rebounding bubble ($R_0\approx1\rm\,mm$, $\Delta p\approx100\rm\,kPa$) moving leftwards while jetting against the dynamic $\del p$, orthogonal to the calculated $p$-contours. (c, d) Bubbles ($R_0\approx2\rm\,mm$, $\Delta p\approx100\rm\,kPa$) rebounding with a jet \emph{towards} a flat rigid surface ($h=5.3\rmax$) and \emph{away} from a flat free surface ($h=5.1\rmax$), respectively. Images were taken using the setup of this letter (a,c,d) and a cavitation tunnel \citep{Avellan1987} (b).}
  	\label{fig_universality}
\end{figure}

Jets produced by cavitation bubbles play a key role in cutting-edge technologies \citep{Mason1996,Dijkink2008,Leighton2010} and erosion \citep{Dear1988,Philipp1998,Ohl2006}. These jets typically arise when a bubble collapses in a liquid of anisotropic pressure: At the ending collapse stage, the bubble surface develops a fast ($\gtrsim\!100\rm\,m\,s^{-1}$ \citep{Benjamin1966,Plesset1971,Dear1988}) liquid jet. This ``micro-jet'' is directed inwards against the local pressure gradient $\del p$ \citep{Katz1999}, defined in the absence of the bubble. While the bubble bounces off its enclosed gas, the micro-jet pierces the bubble and starts penetrating the liquid \citep{Dear1988,Dear1988b,Philipp1998,Ohl2006} unless hitting a boundary. During the regrowth (``rebound'', see \fig{universality}a) of the bubble, the micro-jet becomes visible because of its conical shell of vapor \citep{Chahine1982,Philipp1998,Obreschkow2006,Kobel2009}, here called the ``vapor jet'' (\fig{single_jet}). The velocity and structure of jets of bubbles were modeled and measured in various cases \citep{Plesset1971,Dear1988,Katz1999,Blake1999}, but no general relation is known between the `jet size' and the underlying pressure gradient. Such a relation would allow the modulation of jets by specifically engineering the pressure field; and, vise versa, permit a measurement of a pressure field via static images of jetting bubbles (\fig{universality}).

This letter expands the state-of-the-art in three ways: (i) It presents the first high-speed movies of the jets caused by a gravity-induced pressure gradient $\del p\!=\!\rho\g$ in normal gravity ($g=9.81\rm\,m\,s^{-2}$, liquid density $\rho\approx10^3\rm\,kg\,m^{-3}$). (ii) It performs a systematic study of the vapor jets observed while varying the maximal bubble radius, the liquid viscosity, the liquid pressure, and the pressure gradient. The latter is varied through a modulation of $g$ aboard parabolic flights \footnote{53rd ESA Parabolic Flight Campaign, October 2010}. (iii) A statistical analysis of the data, backed-up by a theoretical derivation, reveals that the jet size scales with a dimensionless jet-parameter $\zeta$.

Our experiment relies on a gravity-induced $\del p$, which exhibits the unique advantage of being uniform in space and time. Such a gradient approximates, to first order, any smooth field $p(\x)=p(0)+\x^\dagger\del p+{\mathcal O}(\partial^2 p)$, where $\del p\equiv\del p(0)$. Examples of $\del p$ are
\begin{subnumcases}
	{\hspace{-6mm}\del p\!=\!\!\label{eq_delp_all}}
	\!\rho\g & \rm gravitational~field, \label{eq_delp_a} \\
	\!-\rho(\v\cdot\!\del)\v~ & \rm stat.~potential~flow,\label{eq_delp_b} \\
	\!+0.2 \rmax\Delta p\,\h/h^3\hspace{-3mm} & \rm rigid~flat~surface, \label{eq_delp_c} \\
	\!-0.2 \rmax\Delta p\,\h/h^3\hspace{-3mm} & \rm free~flat~surface, \label{eq_delp_d}
\end{subnumcases}
where $\v$ is the velocity field, $\rmax$ is the maximal bubble radius before collapse, $\h$ is the shortest vector from the surface to the bubble center, and $\Delta p\equiv p_0-p_{\rm v}$ with $p_0$ being the pressure at cavity level and $p_{\rm v}$ the vapor pressure. \eq{delp_b} follows from momentum conservation of an incompressible, stationary potential flow. Eqs.~(\ref{eq_delp_c}, \ref{eq_delp_d}) provide `effective' time-averages of the self-generated pressure anisotropy of a bubble growing and collapsing close to a flat boundary: $\del p$ is defined such that adding a constant counter-gradient $-\del p$ would suppress the jet according to Ref.~\citep{Blake1988}. Jets described by Eqs.~(\ref{eq_delp_all}a--d) are shown in \fig{universality}. 

\begin{figure}[b]
	\centering 
  	\includegraphics[width=\columnwidth]{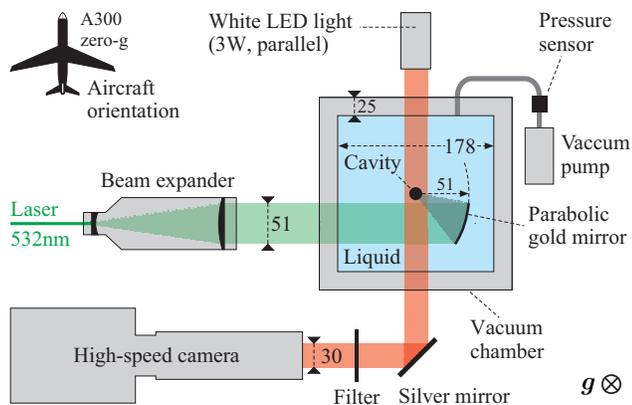}
  	\caption{Schematic view of the experiment flown on parabolic flights. The test-chamber, filled with a water-glycerine mixture of adjustable viscosity, is pressure-controlled. An ultra-spherical cavitation bubble is produced by a $8\rm\,ns$ laser pulse focused with a parabolic mirror at a high convergence angle of $53^\circ$. (Dimensions in mm)}
  	\label{fig_schema}
\end{figure}


\emph{Experimental setup --} Our setup (\fig{schema}) uses a high-speed camera (Photron SA1.1) operating at up to 250,000\,fps with exposure times of $370\rm\,ns$ to record a cavitation bubble generated by a pulsed laser (Quantel CFR 400, 532\,nm, 8\,ns). The laser is focussed inside a liquid volume to form a point-plasma \citep{Byun2004}  ($\rm diameter\lesssim0.1\,mm$), which quickly cools and condensates while growing a bubble that subsequently collapses and rebounds. Whereas past studies \citep{Ohl2006,Philipp1998,
Lauterborn1972} used lenses to focus a laser, we here use for the first time a concave parabolic mirror (\fig{schema}). We found the mirror technique to produce bubbles of much higher sphericity, since reflection is independent of the liquid's refractive index and mirrors allow large angles of convergence (here $53^\circ$) without spherical aberration. Our millimeter-sized bubbles are so spherical that the tiny gravity-induced pressure difference between their top and bottom becomes the dominant source of jet formation. To our knowledge, this experiment provides the first clean movies of gravity-jets in normal gravity conditions. Similar observations in the past \citep{Benjamin1966} required large bubbles ($\rmax\!>\!1\rm\,cm$) in hyper-gravity.

The four controllable experimental parameters are the maximal bubble radius $R_0$ (varied in the range $1-7\rm\,mm$), the liquid pressure at cavity level $p_0$ ($8-80\rm\,kPa$), the norm of the pressure gradient $|\del p|$ ($0-18\rm\,kPa\,m^{-1}$), and the dynamic viscosity $\eta$ ($1-30\rm\,mPa\,s$). These parameters are controlled as follows: A pressure-regulated vacuum pump depressurizes the liquid at a precision of $0.2\rm\,kPa$, while also removing traces of laser-generated gas. The flight manoeuvres (93 ballistic trajectories, straight cruise, 24 steep turns) provide intervals of stable gravity at `0g', `1g', `1.2g', `1.4g', `1.6g', and `1.8g' (i.~e.~$g=1.8\cdot9.81\rm\,m\,s^{-2}$), as well as transition phases, thus offering a wide range of gradients $|\del p|=\rho g$. By adjusting the energy of the laser pulse and the pressure $p_0$, bubbles of various radii $\rmax$ can be obtained. $\rmax$ is then measured at $10\rm\,\mu m$-accuracy on the high-speed movies (e.~g.~Fig.~\ref{fig_universality}a, left). These movies resolve the initial growth and collapse of the bubble into more than 100 frames. Demineralized water is used in the experiments at variable gravity, while ground-based follow-up experiments use water-glycerol mixtures to expand the viscosity range from $\eta=1\rm\,mPa\,s$ (pure water) to $\eta=2\rm\,mPa\,s$ (25\% glycerol mass) and $\eta=30\rm\,mPa\,s$ (75\% glycerol). The addition of glycerol mainly alters $\eta$, but it also affects $p_{\rm v}$, $\rho$, and the sound speed $c$. These variations are accounted for in the data analysis below (Eq.~\ref{eq_def_zeta}).


\emph{Experimental results --} Observations of rebounding bubbles with a gravity-jet are shown in Figs.~\ref{fig_universality}a, \ref{fig_single_jet}a, \ref{fig_single_jet}b, \ref{fig_scaling_law}. Most of these images use a parallel back-light (\fig{schema}), hence the bubbles appear in absorption against a white background. This type of imaging allows a precise measurement of the bubble geometry and a visualization of shocks (Fig.~\ref{fig_universality}a). An alternative front-illumination discloses the internal structure, namely the narrow micro-jet inside the vapor jet (\fig{single_jet}b). The missing jet in `0g' (\fig{universality}a, top) proves the gravitational origin of the jets. Rayleigh-Taylor instabilities during the rebound \citep{Frost1986} can be excluded as jet-drivers because of the torus-topology (\fig{single_jet}b, c).

\begin{figure}[t]
	\centering 
  	\includegraphics[width=\columnwidth]{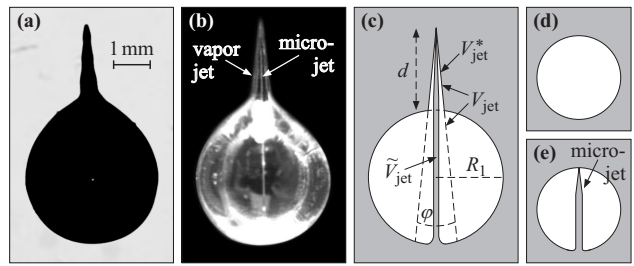}
  	\caption{(a, b) Observations of the gravity-driven jet of a rebounding cavitation bubble ($\rmax=\rm3\,mm$, $\Delta p=10\rm\,kPa$) in normal gravity: (a) using a back-illumination (see \fig{schema}), (b) using a front-illumination and adaptive overlaying of different exposures to increase the dynamic range and sharpness. The vapor jet envelops a narrow micro-jet in agreement with simulations \citep{Blake1999}. (c--e) Model (see text).}
  	\label{fig_single_jet}
\end{figure}

Nearby boundaries can also cause jets by altering the pressure gradient \citep{Chahine1982,Philipp1998,Ohl2006}, and they dominate over gravity if $\lambda\equiv h^2 \rho g/(\rmax\deltap)<0.2$ [square of Eq.~8.8 in Ref.~\citealp{Blake1988}]. Here, $h=55\rm\,mm$ is the distance from the parabolic mirror to the bubble centre. To guarantee accurate results we only retain gravity-dominated cases with $\lambda\geq0.5$, thus keeping a sample of 104 bubbles with jets. Yet, many small bubbles ($\rmax\lesssim\rm2.5~mm$, thus $h/\rmax\gtrsim22$) in `0g' ($\lambda=0$) yield no jet (\fig{universality}a, top), since the influence of boundaries is too weak. Albeit excluded from the analysis, these data is shown in \fig{scaling_law} as a single point.

For each bubble in the sample, each high-speed image of the rebound phase is decomposed into a circular disk and a jet, using a $\chi^2$-fit of a circular top-hat function. The jet volume $\Vvaporjetobs$ (vapor+liquid, \fig{single_jet}c) is then calculated assuming axial symmetry about the jet-axis. Since $\Vvaporjetobs$ only contains a part of the micro-jet, we define an effective volume $\Vvaporjetcor$ as the geometrical extension of $\Vvaporjetobs$ into the bubble (\fig{single_jet}c) -- an approach justified in \textsection~`theoretical model'. The relation between $\Vvaporjetcor$ and $\Vvaporjetobs$ depends on the cone angle $\varphi$. This angle is measured along the edges of the vapor jet rather than at its tip to bypass potential deformations of the tip by surface tension. When $\Vvaporjetobs$ is maximal, we observe $\varphi\approx4^\circ$ across all bubbles. Trigonometry then implies \footnote{Define $z^3\equiv(\pi/3)\tan^2(\varphi/2)$ $\Rightarrow$ $\Vvaporjetobs=z^3d^3$ and $\Vvaporjetcor=z^3(d\!+\!2R_1)^3=(zd\!+\!2zR_1)^3\approx(\Vvaporjetobs^{\!\!1/3}\!+\!0.2R_1)^3$.} $\Vvaporjetcor^{\rm max} = (\Vvaporjetobsmax^{1/3}+0.2\rreb)^3$ where $\rreb$ is the maximal bubble radius during the rebound. We finally define a ``normalized jet volume'' as
\beq\label{eq_def_ejet}
	\ejet\,\equiv\,\Vvaporjetcor^{\rm max}/[(4\pi/3)\rreb^3]\,.
\eeq

Aiming for a model of $\ejet$, we adopt the Ansatz that $\ejet$ is proportional to a non-dimensional parameter $\zeta$, defined as a power law of the parameters $\rmax$, $\rho$, $g$, $p_0$, $\eta$, and the liquid compressibility (sound speed $c$). The surface tension $\sigma$ is neglected as justified for our relatively large bubbles \footnote{The ``Rayleigh pressure'' $\Delta p+\frac{3}{2}\rho\dot{R}^2(t)$ exceeds the ``surface pressure'' $2\sigma/R(t)$ by a factor $>\!10^2$ at all times $t\!\in\![0,T_{\rm c}]$ for bubbles of $\rmax>0.1\rm\,mm$ at $p_0=1\rm\,bar$ and $\rmax\!>\!1\rm\,mm$ at $p_0=0.1\rm\,bar$. Strictly, this argument is restricted to a spherical collapse; see \textsection~`theoretical model' for an extension to jets.}, but we will show (theory below) that even the jets of much smaller bubbles remain insignificantly affected by surface tension. The most general non-dimensional form of $\zeta$ then reads
\beq\label{eq_def_zeta}
	\zeta = \left(\rmax^\alpha\,\rho^\beta\,g~\deltap^{\alpha-\beta-1}\,c^{-\alpha+2\beta-1}\,\eta^{-\alpha+1}\right)^\gamma,
\eeq
where $\alpha$, $\beta$, $\gamma$ are free parameters. To determine $\alpha$, $\beta$, $\gamma$ we perform a $\chi^2$-fit, minimizing the uncertainty-weighted rms of $\ejet/\zeta$ over the 104 data points. This yields $\alpha=1.04\pm0.03$, $\beta=1.05\pm0.20$, $\gamma=0.98\pm0.10$ (with $\chi^2=0.9$), where the ranges are $67\%$ confidence intervals obtained by bootstrapping the data \citep{Efron1987}. Since $\alpha=\beta=\gamma=1$ is consistent with the data, it seems natural to adopt this choice. Substituting $|\del p|=\rho g$ then reduces \eq{def_zeta} to
\beq\label{eq_general_scaling_law}
	\zeta = |\del p|\,\rmax/\deltap\,.
\eeq
\fig{scaling_law} shows the measured values of $\ejet$ as a function of $\zeta$ together with the linear regression (solid line)
\beq\label{eq_scaling_law}
	\ejet = 5.4\,\zeta\,.
\eeq
A remarkable feature of this proportionality relation (Eqs.~\ref{eq_general_scaling_law}, \ref{eq_scaling_law}) is its independence of the viscosity $\eta$, as verified for $\eta$-variations by a factor 30 (\fig{scaling_law}).

If no vapor jet is observed ($\Vvaporjetobsmax=0$), Eq.~(\ref{eq_def_ejet}) implies $\ejet=\ejet^{\rm min}\approx0.002$. The inequality $\ejet^{\rm min}>0$ reflects that when no vapor jet forms, a micro-jet may still be present. In fact, no vapor jet arises in all cases between no micro-jet (\fig{single_jet}d) and a micro-jet that just touches the bubble surface (\fig{single_jet}e). This range of indeterminacy is shaded in \fig{scaling_law}. Using Eq.~(\ref{eq_scaling_law}) the threshold value $\zeta_{\rm c}$, where $\ejet(\zeta_{\rm c})=\ejet^{\rm min}$, reads $\zeta_{\rm c}\approx4\cdot10^{-4}$. The jet only pierces the bubble surface if $\zeta>\zeta_{\rm c}$, i.~e.~$\zeta_{\rm c}$ delimits the topological transition between a sphere and a torus. To confirm this conjecture further investigations of this transition are needed.

\begin{figure}[t]
	\centering 
  	\includegraphics[width=\columnwidth]{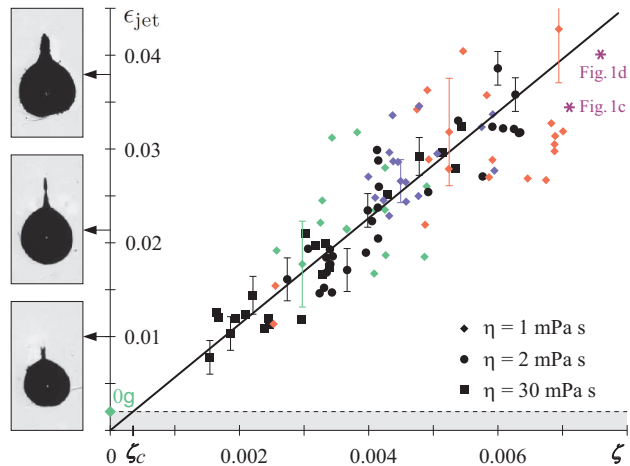}
  	\caption{Scaling law. Black points are data at varying $\rmax$, $p_0$, $\eta$ and fixed $g=9.81\rm\,m\,s^{-2}$. Colored points are data at varying $\rmax$, $p_0$, $g$ and fixed $\eta=1\rm\,mPa\,s$ (green: $g\!<\!11\rm\,m\,s^{-2}$, red: $g\!>\!16\rm\,m\,s^{-2}$, blue: intermediate). Some 67\% measurement uncertainties are shown by the error bars. The solid line is the weighted regression (\eqp{scaling_law}). The zone of experimental indeterminacy, covering the cases between \fig{single_jet}d and \fig{single_jet}e, is grey-shaded.}
  	\label{fig_scaling_law}
\end{figure}


\emph{Theoretical model --} The relation $\ejet\propto\zeta$ will now be derived from first principles. By conservation of momentum, the micro-jet momentum equals the integrated momentum accumulated by the liquid during the bubble growth and collapse. This momentum, called ``Kelvin impulse'', was explored by Blake \citep{Blake1988}. It can be computed as $\I=\int_{-T_{\rm c}}^{T_{\rm c}}\d t\int_{S(t)}\d\F$, where $T_{\rm c}$ is the collapse time, $S$ is the bubble surface (here assumed spherical), and $\d\F=-p\,\d\S$ is the force acting on the bubble surface. Spherically symmetric (isotropic) terms in the pressure field $p$ vanish in the integral over $S$. Hence, we only consider anisotropic pressure terms, here given by a constant $\del p$, defined as the pressure gradient in the absence of a bubble. The bubble adds an additional gradient that varies with the normalized time $\tau\equiv t/T_{\rm c}$. This bubble-generated gradient contains a radial, spherically symmetric term (\textsection 3.2.3 in \citep{Franc2004}) vanishing in the momentum integral, and a linear term proportional to $\del p$ caused by the motion of the bubble center. Thus, neglecting isotropic terms, $\d\F\!=\!-f(\tau)(\del p\cdot\R)\,\d\S$, where $f(\tau)$ is a scalar function and $\R(t)$ is the vector from the bubble center to a surface element. In this generic model, the micro-jet momentum $\I$ solves to \footnote{Use spherical coord.~$(R,\phi,\theta)$ with $\theta\equiv\angle(\del p,\R)$ $\Rightarrow$ $\del p\cdot\R\!=\!|\del p|R\cos\theta$. Only the $z$-projection of $\d\S$, $\d S_{\rm z}\!=\!\d\phi\,\d\theta R^2\sin\theta\cos\theta$, persists in the $\phi$-integral $\Rightarrow$ $\int_{S(t)}\d\F\!=\!-f R^3\del p\int_0^{2\pi}\!\d\phi\int_0^{\pi}\sin\theta\,\cos^2\!\theta\,\d\theta$ $\Rightarrow$ $\I\!=\!-(4\pi/3)\del p\int_{-T_{\rm c}}^{T_{\rm c}}f R^3\d t$. According to Rayleigh $R(t)\!=\!\rmax\tilde{R}(\tau)$, where $\tilde{R}(\tau)$ is a unique function, as is $f\!=\!f(\tau)$ $\Rightarrow$ $\I\!=\!-(4\pi/3)\del p\rmax^3 T_{\rm c}\int_{-1}^1f(\tau)\tilde{R}(\tau)^3\d t\propto-\del p\rmax^3T_{\rm c}$.}
\beq\label{eq_kelvin_impulse}
	\I \propto -\del p\rmax^3 T_{\rm c}~.
\eeq
$T_{\rm c}$ is the key term, where side-effects can intervene. While the Rayleigh-theory \citep{Rayleigh1917} implies $T_{\rm c}\!\approx\!0.915\rmax\sqrt{\rho/\deltap}$, the Plesset-theory \citep{Plesset1977} details that: (i) Incondensable gas increases $T_{\rm c}$ (see \textsection~`Discussion'). (ii) Surface tension decreases $T_{\rm c}$ -- by less than $1\%$ for our bubbles and by about $7\%$ for a micro-bubble ($\rmax=10\rm\,\mu m$) in water at standard conditions ($\sigma=0.07\rm\,N\,m^{-1}$, $p_0=10^5\rm\,Pa$). Carried along to \eq{mjet}, this $7\%$-effect reduces the jet mass by only $14\%$, comparable to a $1\rm\,\mu m$-measurement error of $\rmax$. (iii) Viscosity increases $T_{\rm c}$ for small bubbles, but the effect is even weaker -- about 1\% for a bubble with $\rmax=10\rm\,\mu m$ (at $\eta=1\rm\,mPa\,s$, $p_0=10^5\rm\,Pa$). Hereafter $T_{\rm c}$ in \eq{kelvin_impulse} will hence be approximated as $0.915\rmax\sqrt{\rho/\deltap}$.

In analogy to the ``Kelvin impulse'' we now introduce a kinetic ``Kelvin energy'' $E=2\int_0^{\rmax}\int_{S(t)}|\d\F\cdot\d\R|$, resulting from the work done by the same anisotropic forces that generate the jet's momentum. Thus \footnote{Use spherical coord.~as above. Thus, $\d\F\cdot\d\R=\d R\,\d\phi\,\d\theta\,|\del p|f R^3\sin\theta\cos\theta$ $\Rightarrow$ $\int_{S(t)}|\d\F\cdot\d\R|=2\pi|\del p|f R^3\d R$. Since $f$ can be rewritten as a function of $R/\rmax$, $\int_0^{\rmax}f R^3\d R\propto\rmax^4$ $\Rightarrow$ \eq{kelvin_energy}.},
\beq\label{eq_kelvin_energy}
	E \propto |\del p|\,\rmax^4~.
\eeq

Defining $m$ and $\v$ as the mass and spatially averaged velocity of the fully developed micro-jet (Fig.~\ref{fig_single_jet}) implies $\I=m\v$ and $E\propto m\v^2$. Eqs.~(\ref{eq_kelvin_impulse}, \ref{eq_kelvin_energy}) then yield
\begin{subequations}
	\begin{align}
		\v\, & \propto -\sqrt{\Delta p/\rho\,}~\e\,, \label{eq_vjet} \\
		m & \propto \,|\del p|\,\rmax^4\,\rho/\Delta p\,, \label{eq_mjet}
	\end{align}
\end{subequations}
where $\e\equiv\del p/|\del p|$. \eq{vjet} is a known relation \citep{Plesset1971}, while \eq{mjet} is of interest regarding the jet volume.
We hypothesize that the effective jet volume scales with the micro-jet volume, $\Vvaporjetcor^{\rm max}=\varepsilon\, m\,\rho^{-1}$, at an efficiency $\varepsilon\propto\rreb^3/\rmax^3$. This approximation derives from the observation that the vapor jet grows out of the rebound bubble, thus consuming a fraction of the rebound volume $\propto R_1^3$. Eqs.~(\ref{eq_def_ejet}, \ref{eq_mjet}) then imply
\beq\label{eq_theoretical_scaling_law}
	\ejet\propto|\del p|\,\rmax/\deltap\equiv\zeta\,.
\eeq
Alternatively, \eq{theoretical_scaling_law} also results from assuming $\ejet$ proportional to $E/E_0$, where $E_0=(4\pi/3)R_0^3\Delta p$.


\emph{Discussion --} In summary, Eqs.~(\ref{eq_scaling_law}, \ref{eq_theoretical_scaling_law}) demonstrate experimentally and theoretically that the normalized jet volume $\ejet$ is proportional to $\zeta$. Along this discovery subtle issues were encountered that are worth explaining: (i) \eq{theoretical_scaling_law} is not `scale invariant' as it depends on $\rmax$: given $\del p$, big bubbles yield larger $\ejet$ than small ones. Formal scale invariance is nonetheless recovered using an adequately normalized gradient $\tilde{\del}\equiv\sum_{\rm i=1}^3\partial/(\partial x_{\rm i}/\rmax)\e_{\rm i}$ $\Rightarrow$ $\zeta=|\tilde{\del}p|/\Delta p$. (ii) How can a micro-jet survive inside the hot gas \citep{Baghdassarian2001} during the collapse point? This feature might be attributed to the life-time of the hot gas ($<1\rm\,\mu s$, \citep{Baghdassarian2001}) being too short to evaporate the jet. (iii) We neglected incondensable gas inside the bubble, since the measured bubble radii $R(t)$ agree within 1\% with the Rayleigh-equation \citep{Rayleigh1917}, when neglecting incondensable gas. If bubbles contain significant amounts of incondensable gas, they do not fully collapse, hence less concentrating their Kelvin-impulse. (iv) While Eqs.~(\ref{eq_scaling_law}, \ref{eq_theoretical_scaling_law}) rely on stationary $\del p$'s, they also approximate non-stationary situations, if the characteristic time-scales are comparable to or above $T_{\rm c}$. This is illustrated by the data of Figs.~\ref{fig_universality}c, d (where $\ejet=0.2\,\rmax^2\,h^{-2}$) plotted as stars in Fig.~\ref{fig_scaling_law}. (v) We here considered jet parameters $\zeta<0.008$. Larger $\zeta$ may produce more complex bubble-jet morphologies, requiring a decomposition into spherical harmonics.


Supported by the European Space Agency ESA and the Swiss NSF (200020-116641, PBELP2-130895).


\begin{thebibliography}{24}
\expandafter\ifx\csname natexlab\endcsname\relax\def\natexlab#1{#1}\fi
\expandafter\ifx\csname bibnamefont\endcsname\relax
  \def\bibnamefont#1{#1}\fi
\expandafter\ifx\csname bibfnamefont\endcsname\relax
  \def\bibfnamefont#1{#1}\fi
\expandafter\ifx\csname citenamefont\endcsname\relax
  \def\citenamefont#1{#1}\fi
\expandafter\ifx\csname url\endcsname\relax
  \def\url#1{\texttt{#1}}\fi
\expandafter\ifx\csname urlprefix\endcsname\relax\def\urlprefix{URL }\fi
\providecommand{\bibinfo}[2]{#2}
\providecommand{\eprint}[2][]{\url{#2}}

\bibitem[{\citenamefont{Mason et~al.}(1996)\citenamefont{Mason, Paniwnyk, and
  Lorimer}}]{Mason1996}
\bibinfo{author}{\bibfnamefont{T.~J.} \bibnamefont{Mason}},
  \bibinfo{author}{\bibfnamefont{L.}~\bibnamefont{Paniwnyk}}, \bibnamefont{and}
  \bibinfo{author}{\bibfnamefont{J.~P.} \bibnamefont{Lorimer}},
  \bibinfo{journal}{Ultrasonics Sonochemistry} \textbf{\bibinfo{volume}{3}},
  \bibinfo{pages}{253} (\bibinfo{year}{1996}).

\bibitem[{\citenamefont{Dijkink and Ohl}({2008})}]{Dijkink2008}
\bibinfo{author}{\bibfnamefont{R.}~\bibnamefont{Dijkink}} \bibnamefont{and}
  \bibinfo{author}{\bibfnamefont{C.-D.} \bibnamefont{Ohl}},
  \bibinfo{journal}{{Lab on a Chip}} \textbf{\bibinfo{volume}{{8}}},
  \bibinfo{pages}{1676} (\bibinfo{year}{{2008}}).

\bibitem[{\citenamefont{Leighton and Cleveland}(2010)}]{Leighton2010}
\bibinfo{author}{\bibfnamefont{T.~G.} \bibnamefont{Leighton}} \bibnamefont{and}
  \bibinfo{author}{\bibfnamefont{R.~O.} \bibnamefont{Cleveland}},
  \bibinfo{journal}{Proc. Inst. Mech. Eng. H} \textbf{\bibinfo{volume}{224}},
  \bibinfo{pages}{317} (\bibinfo{year}{2010}).

\bibitem[{\citenamefont{Dear and Field}(1988)}]{Dear1988}
\bibinfo{author}{\bibfnamefont{J.~P.} \bibnamefont{Dear}} \bibnamefont{and}
  \bibinfo{author}{\bibfnamefont{J.~E.} \bibnamefont{Field}},
  \bibinfo{journal}{J. Fluid Mech.} \textbf{\bibinfo{volume}{190}},
  \bibinfo{pages}{409} (\bibinfo{year}{1988}).

\bibitem[{\citenamefont{{Philipp} and {Lauterborn}}({1998})}]{Philipp1998}
\bibinfo{author}{\bibfnamefont{A.}~\bibnamefont{{Philipp}}} \bibnamefont{and}
  \bibinfo{author}{\bibfnamefont{W.}~\bibnamefont{{Lauterborn}}},
  \bibinfo{journal}{{J. Fluid Mech.}} \textbf{\bibinfo{volume}{{361}}},
  \bibinfo{pages}{75} (\bibinfo{year}{{1998}}).

\bibitem[{\citenamefont{Ohl et~al.}(2006)\citenamefont{Ohl, Arora, Dijkink,
  Janve, and Lohse}}]{Ohl2006}
\bibinfo{author}{\bibfnamefont{C.-D.} \bibnamefont{Ohl}},
  \bibinfo{author}{\bibfnamefont{M.}~\bibnamefont{Arora}},
  \bibinfo{author}{\bibfnamefont{R.}~\bibnamefont{Dijkink}},
  \bibinfo{author}{\bibfnamefont{V.}~\bibnamefont{Janve}}, \bibnamefont{and}
  \bibinfo{author}{\bibfnamefont{D.}~\bibnamefont{Lohse}},
  \bibinfo{journal}{Applied Physics Letters} \textbf{\bibinfo{volume}{89}},
  \bibinfo{pages}{074102} (\bibinfo{year}{2006}).

\bibitem[{\citenamefont{Benjamin and Ellis}(1966)}]{Benjamin1966}
\bibinfo{author}{\bibfnamefont{T.~B.} \bibnamefont{Benjamin}} \bibnamefont{and}
  \bibinfo{author}{\bibfnamefont{A.~T.} \bibnamefont{Ellis}},
  \bibinfo{journal}{Phil. Trans. R. Soc. Lond. A}
  \textbf{\bibinfo{volume}{260}}, \bibinfo{pages}{221} (\bibinfo{year}{1966}).

\bibitem[{\citenamefont{{Plesset} and {Chapman}}(1971)}]{Plesset1971}
\bibinfo{author}{\bibfnamefont{M.}~\bibnamefont{{Plesset}}} \bibnamefont{and}
  \bibinfo{author}{\bibfnamefont{R.~B.} \bibnamefont{{Chapman}}},
  \bibinfo{journal}{J. Fluid Mech.} \textbf{\bibinfo{volume}{47}},
  \bibinfo{pages}{283} (\bibinfo{year}{1971}).

\bibitem[{\citenamefont{Katz}(1999)}]{Katz1999}
\bibinfo{author}{\bibfnamefont{J.~I.} \bibnamefont{Katz}},
  \bibinfo{journal}{Proc. R. Soc. Lond. A} \textbf{\bibinfo{volume}{455}},
  \bibinfo{pages}{323} (\bibinfo{year}{1999}).

\bibitem[{\citenamefont{Dear et~al.}(1988)\citenamefont{Dear, Field, and
  Walton}}]{Dear1988b}
\bibinfo{author}{\bibfnamefont{J.~P.} \bibnamefont{Dear}},
  \bibinfo{author}{\bibfnamefont{J.~E.} \bibnamefont{Field}}, \bibnamefont{and}
  \bibinfo{author}{\bibfnamefont{A.~J.} \bibnamefont{Walton}},
  \bibinfo{journal}{Nature} \textbf{\bibinfo{volume}{332}},
  \bibinfo{pages}{505} (\bibinfo{year}{1988}).

\bibitem[{\citenamefont{Chahine}({1982})}]{Chahine1982}
\bibinfo{author}{\bibfnamefont{G.~L.} \bibnamefont{Chahine}},
  \bibinfo{journal}{{Appl. Sci. Res.}} \textbf{\bibinfo{volume}{{38}}},
  \bibinfo{pages}{187} (\bibinfo{year}{{1982}}).

\bibitem[{\citenamefont{{Obreschkow} et~al.}(2006)\citenamefont{{Obreschkow},
  {Kobel}, {Dorsaz}, {de Bosset}, {Nicollier}, and {Farhat}}}]{Obreschkow2006}
\bibinfo{author}{\bibfnamefont{D.}~\bibnamefont{{Obreschkow}}},
  \bibinfo{author}{\bibfnamefont{P.}~\bibnamefont{{Kobel}}},
  \bibinfo{author}{\bibfnamefont{N.}~\bibnamefont{{Dorsaz}}},
  \bibinfo{author}{\bibfnamefont{A.}~\bibnamefont{{de Bosset}}},
  \bibinfo{author}{\bibfnamefont{C.}~\bibnamefont{{Nicollier}}},
  \bibnamefont{and} \bibinfo{author}{\bibfnamefont{M.}~\bibnamefont{{Farhat}}},
  \bibinfo{journal}{Phys.~Rev.~Lett.} \textbf{\bibinfo{volume}{97}},
  \bibinfo{pages}{094502} (\bibinfo{year}{2006}).

\bibitem[{\citenamefont{{Kobel} et~al.}(2009)\citenamefont{{Kobel},
  {Obreschkow}, {Dorsaz}, {De Bosset}, and {Farhat}}}]{Kobel2009}
\bibinfo{author}{\bibfnamefont{P.}~\bibnamefont{{Kobel}}},
  \bibinfo{author}{\bibfnamefont{D.}~\bibnamefont{{Obreschkow}}},
  \bibinfo{author}{\bibfnamefont{N.}~\bibnamefont{{Dorsaz}}},
  \bibinfo{author}{\bibfnamefont{A.}~\bibnamefont{{De Bosset}}},
  \bibnamefont{and} \bibinfo{author}{\bibfnamefont{M.}~\bibnamefont{{Farhat}}},
  \bibinfo{journal}{Experiments in Fluids} \textbf{\bibinfo{volume}{47}},
  \bibinfo{pages}{39} (\bibinfo{year}{2009}).

\bibitem[{\citenamefont{{Blake} et~al.}(1999)\citenamefont{{Blake}, {Keen},
  {Tong}, and {Wilson}}}]{Blake1999}
\bibinfo{author}{\bibfnamefont{J.~R.} \bibnamefont{{Blake}}},
  \bibinfo{author}{\bibfnamefont{G.~S.} \bibnamefont{{Keen}}},
  \bibinfo{author}{\bibfnamefont{R.~P.} \bibnamefont{{Tong}}},
  \bibnamefont{and} \bibinfo{author}{\bibfnamefont{M.}~\bibnamefont{{Wilson}}},
  \bibinfo{journal}{Philos T R Soc A} \textbf{\bibinfo{volume}{357}},
  \bibinfo{pages}{251} (\bibinfo{year}{1999}).

\bibitem[{\citenamefont{{Avellan} et~al.}(1987)\citenamefont{{Avellan},
  {Henry}, and {Rhyming}}}]{Avellan1987}
\bibinfo{author}{\bibfnamefont{F.}~\bibnamefont{{Avellan}}},
  \bibinfo{author}{\bibfnamefont{P.}~\bibnamefont{{Henry}}}, \bibnamefont{and}
  \bibinfo{author}{\bibfnamefont{I.}~\bibnamefont{{Rhyming}}},
  \bibinfo{journal}{Proc. Int. Symp. on Cav. Research Facilities and Tech.}
  \textbf{\bibinfo{volume}{57}}, \bibinfo{pages}{49} (\bibinfo{year}{1987}).

\bibitem[{\citenamefont{Blake}(1988)}]{Blake1988}
\bibinfo{author}{\bibfnamefont{J.~R.} \bibnamefont{Blake}},
  \bibinfo{journal}{J. Austr. Math. Soc. B} \textbf{\bibinfo{volume}{30}},
  \bibinfo{pages}{127} (\bibinfo{year}{1988}).

\bibitem[{\citenamefont{{Byun} and {Kwak}}(2004)}]{Byun2004}
\bibinfo{author}{\bibfnamefont{K.-T.} \bibnamefont{{Byun}}} \bibnamefont{and}
  \bibinfo{author}{\bibfnamefont{H.-Y.} \bibnamefont{{Kwak}}},
  \bibinfo{journal}{Japanese Journal of Applied Physics}
  \textbf{\bibinfo{volume}{43}}, \bibinfo{pages}{621} (\bibinfo{year}{2004}).

\bibitem[{\citenamefont{{Lauterborn}}(1972)}]{Lauterborn1972}
\bibinfo{author}{\bibfnamefont{W.}~\bibnamefont{{Lauterborn}}},
  \bibinfo{journal}{Applied Physics Letters} \textbf{\bibinfo{volume}{21}},
  \bibinfo{pages}{27} (\bibinfo{year}{1972}).

\bibitem[{\citenamefont{Frost and Sturtevant}({1986})}]{Frost1986}
\bibinfo{author}{\bibfnamefont{D.}~\bibnamefont{Frost}} \bibnamefont{and}
  \bibinfo{author}{\bibfnamefont{B.}~\bibnamefont{Sturtevant}},
  \bibinfo{journal}{{J. Heat Transfer-Trans. ASME}}
  \textbf{\bibinfo{volume}{{108}}}, \bibinfo{pages}{418}
  (\bibinfo{year}{{1986}}).

\bibitem[{\citenamefont{{Efron}}(1987)}]{Efron1987}
\bibinfo{author}{\bibfnamefont{B.}~\bibnamefont{{Efron}}}, \bibinfo{journal}{J.
  Am. Stat. Assoc.} \textbf{\bibinfo{volume}{82}}, \bibinfo{pages}{171}
  (\bibinfo{year}{1987}).

\bibitem[{\citenamefont{Franc and Michel}(2004)}]{Franc2004}
\bibinfo{author}{\bibfnamefont{J.-P.} \bibnamefont{Franc}} \bibnamefont{and}
  \bibinfo{author}{\bibfnamefont{J.-M.} \bibnamefont{Michel}},
  \emph{\bibinfo{title}{Fundamentals of Cavitation}}
  (\bibinfo{publisher}{Kluwer Academic Publishers}, \bibinfo{year}{2004}).

\bibitem[{\citenamefont{Rayleigh}(1917)}]{Rayleigh1917}
\bibinfo{author}{\bibfnamefont{L.}~\bibnamefont{Rayleigh}},
  \bibinfo{journal}{Phil. Mag.} \textbf{\bibinfo{volume}{34}},
  \bibinfo{pages}{94} (\bibinfo{year}{1917}).

\bibitem[{\citenamefont{{Plesset} and {Prosperetti}}(1977)}]{Plesset1977}
\bibinfo{author}{\bibfnamefont{M.~S.} \bibnamefont{{Plesset}}}
  \bibnamefont{and}
  \bibinfo{author}{\bibfnamefont{A.}~\bibnamefont{{Prosperetti}}},
  \bibinfo{journal}{Annual Review of Fluid Mechanics}
  \textbf{\bibinfo{volume}{9}}, \bibinfo{pages}{145} (\bibinfo{year}{1977}).

\bibitem[{\citenamefont{Baghdassarian et~al.}(2001)\citenamefont{Baghdassarian,
  Chu, Tabbert, and Williams}}]{Baghdassarian2001}
\bibinfo{author}{\bibfnamefont{O.}~\bibnamefont{Baghdassarian}},
  \bibinfo{author}{\bibfnamefont{H.-C.} \bibnamefont{Chu}},
  \bibinfo{author}{\bibfnamefont{B.}~\bibnamefont{Tabbert}}, \bibnamefont{and}
  \bibinfo{author}{\bibfnamefont{G.~A.} \bibnamefont{Williams}},
  \bibinfo{journal}{Phys. Rev. Lett.} \textbf{\bibinfo{volume}{86}},
  \bibinfo{pages}{4934} (\bibinfo{year}{2001}).

\end{thebibliography}

\end{document}